# EFFICIENCY OF FEEDBACKS FOR SUPPRESSION OF TRANSVERSE INSTABILITIES OF BUNCHED BEAMS

Alexey Burov

Fermilab, Batavia, IL 60510

*Abstract*

Which gain and phase have to be set for a bunch-by-bunch transverse damper, and at which chromaticity it is better to stay? These questions are considered for three models: the two-particle model with possible quadrupole wake, the author's Nested Head-Tail (NHT) model with the broadband impedance, and the NHT with the LHC impedance model.

## INTRODUCTION

How one has to use the bunch-by-bunch damper for the most efficient suppression of the transverse instabilities of bunched beams? This problem has so many parameters and input functions, associated with the beam, impedance and damper, that its full treatment seems hardly possible. In this situation, analytical studies of especially interesting cases by means of available models suggest a reasonable way to get a better understanding. This paper is an example of that sort of research.

First, the simplest of all, the two-particle model [Kohaupt, Talman, Ruth], is explored, with constant dipole and quadrupole wakes, chromaticity and feedback. For all cases considered in this paper, the feedback is assumed to see only the bunch centroid, and to respond by kicking the bunch as a whole. In other words, the damper is assumed to be flat within the bunch length. After the two-particle model, the next in complexity is the hollow-beam or air-bag one [Chao], which represents the bunch by a circle in the longitudinal phase space. In this paper, though, we skip that model, jumping directly to its generalization developed by the author, the Nested Head-Tail (NHT), which represents the bunch by any number of concentric air-bags [NHT] and takes into account intra- and inter-bunch wake fields, as well as the damper. Although NHT allows computing Landau damping, that sort of analysis is predominantly left beyond the scope of this paper, except for some general considerations in the last section.

With the NHT, two impedance models were analyzed: the broadband impedance and the LHC model [Nicolas]. For both of them, it is shown that with the resistive damper there is an area of stability in the gain-chromaticity plane, centered at slightly negative chromaticity, where the multi-bunch beam is stable even without radiation or Landau damping. It is shown that the shapes of these areas of stability, as well as their limitations by the beam intensity, vary a lot. While for the broadband impedance this area allows to increase the TMCI threshold by up to a factor of four, for the LHC model it disappears almost exactly at the same intensity as the no-gain, zero-chromaticity TMCI onset (addressed below just as the TMCI threshold), so one cannot use it close to or above this threshold. That is why at sufficiently high intensity of separated LHC beams, the optimal strategy is to work at high chromaticity and sufficient gain area, or in the *valley of slow instabilities*.

## TWO-PARTICLE MODEL

### Round Chamber

The two-particle model at constant wake, zero chromaticity and round vacuum chamber has been studied analytically by R. Ruth and S. Myers in the 80s [Ruth, Myers, Chao]. In this section we are going to reproduce their results with some more details.

Assuming betatron frequency $\omega_b$ to be high compared with other frequencies and rates, two second-order differential equations for the offsets $x_{1,2} = a_{1,2} e^{-i\omega_b t}$ reduce to a couple of first-order differential equations on the slow amplitudes $a_{1,2}$:

$$\dot{a}_1 = -ig(a_1 + a_2) + iwa_2 \Theta(t - T/2);$$
$$\dot{a}_2 = -ig(a_1 + a_2) + iwa_1 \Theta(T/2 - t); \quad (1)$$
$$\omega_b \gg \omega_s = 2\pi/T, g, w;$$

Here $g$ and $w$ are the normalized damper gain and the wake value, $t$ is time, $T$ is the synchrotron period, $\Theta(t)$ is the Heaviside theta-function; the equations are applied for $0 \leq t \leq T$. Complex values for the gain and real values for the wake are assumed. We will analyze stability of Eq.(1) for the real and imaginary gains independently: in the former case, the damper is called reactive, while in the latter it is resistive. For all the cases, the problem is solved by means of constructing the transfer matrix $\mathbf{M}$, which maps the initial amplitudes onto their values after the synchrotron period:

$$\mathbf{A}(T) = \mathbf{M} \cdot \mathbf{A}(0); \quad \mathbf{A} = (a_1, a_2)^T. \quad (2)$$

For the equations with piecewise constant coefficients, this entire transfer matrix is just a product of partial transfers. Doing these computations, the determinant and trace of the transfer matrix are found as:

$$\operatorname{Det} \mathbf{M} = e^{-2igT};$$
$$\operatorname{Tr} \mathbf{M} = \frac{(2g-w)^2 \cos\left(\sqrt{g(g-w)}T\right) - w^2}{2g(g-w)} e^{-igT}. \quad (3)$$

From here, the eigenvalues of the 2⊗2 matrix $\mathbf{M}$ can be obtained from a quadratic equation:
$$\lambda_{1,2} = \frac{\operatorname{Tr} \mathbf{M}}{2} \pm \sqrt{\frac{(\operatorname{Tr} \mathbf{M})^2}{4} - \operatorname{Det} \mathbf{M}}. \quad (4)$$

The instability growth rate then follows:
$$\tau^{-1} = \max_k \left(\ln |\lambda_k|\right)/T. \quad (5)$$

For reactive dampers, with real $g$, the stability condition $\tau^{-1} \leq 0$ can be presented as $\left|\operatorname{Tr} \mathbf{M} / \sqrt{\operatorname{Det} \mathbf{M}}\right| \leq 2$, or [Ruth, MathErr]:
$$\left(\sqrt{\frac{g}{g-w}} + \sqrt{\frac{g-w}{w}}\right)^2 \sin^2\left(\sqrt{g(g-w)}T/2\right) \leq 4 \quad (6)$$

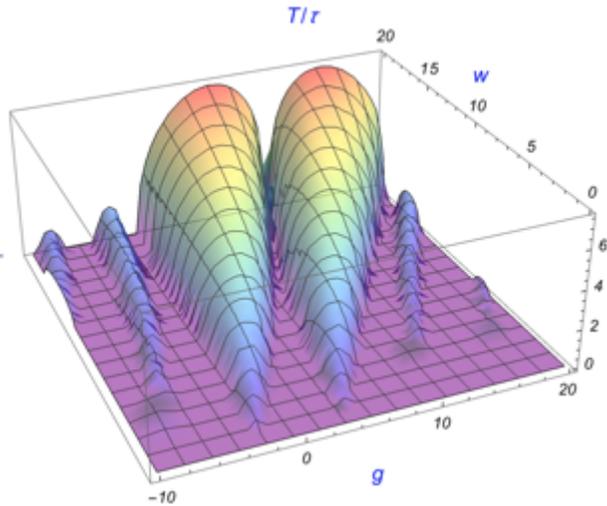

Fig. 1: Two-particle growth rate versus gain $g$ and constant wake value $w$ for reactive damper and zero chromaticity. All the values are in the units of the inverse synchrotron period $1/T$.

Plots for the growth rate for reactive and resistive dampers are presented in Figs. 1-3.

Looking at these results, it could be concluded that the reactive damper is more efficient than resistive. While in the former case the instability threshold can be increased many times, in the latter the intensity threshold is zero for non-zero gains, which is seen in Fig.3. However, S. Myers pointed out [Myers] that at PEP experiment "at high gain the resistive feedback produced a larger enhancement of the threshold current than the reactive feedback. This is in contradiction to the results obtained from theory and is not fully understood."

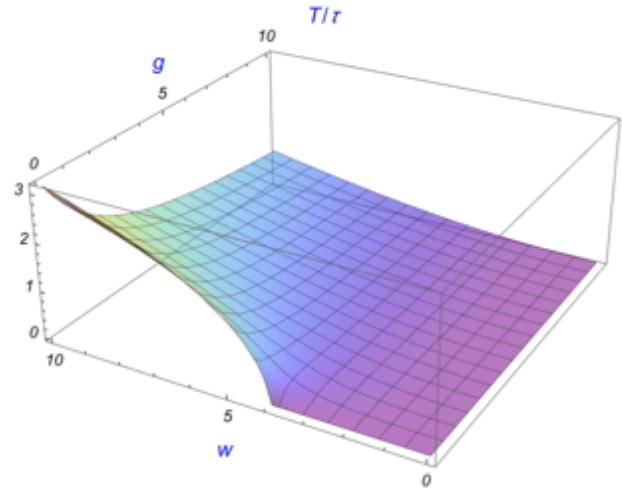

Fig. 2: The same as Fig. 1, but for the resistive damper

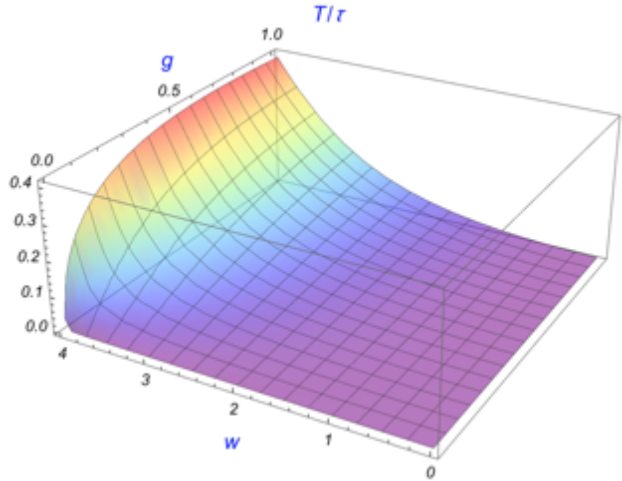

Fig. 3: Low gain, low wake area for the resistive damper.

We should not forget, however, the extreme simplicity of the two-particle model at zero chromaticity. Below, when we will take into account both the chromaticity and the variety of head-tail modes, this contradiction disappears.

*Flat Chamber*

For the vertical instability and the resistive wall Yokoya factor, Eqs.(1) are modified as
$$\dot{a}_1 = -ig(a_1 + a_2) + iw(a_2 + a_1/2)\Theta(t - T/2);$$
$$\dot{a}_2 = -ig(a_1 + a_2) + iw(a_1 + a_2/2)\Theta(T/2 - t). \quad (7)$$

The determinant and trace of the transfer matrix are computed as following:

$$\text{Det}\,\mathbf{M} = e^{-2igT + iwT/2};$$

$$\text{Tr}\,\mathbf{M} = \frac{8(2g-w)^2 \cos\left(T\sqrt{g^2 - gw + w^2/16}\right) - 6w^2}{16g^2 - 16gw + w^2} e^{-igT + iwT/4}.$$

Growth rate plots for the reactive and resistive dampers are barely distinguishable from the similar cases of the round chamber, so that there is no reason to present them; the TMCI threshold is only a few percent higher.

For the horizontal degree of freedom, the equations of motion are written as

$$\dot{a}_1 = -ig(a_1 + a_2) + i\frac{w}{2}(a_2 - a_1)\Theta(t - T/2); \quad (8)$$

$$\dot{a}_2 = -ig(a_1 + a_2) + i\frac{w}{2}(a_1 - a_2)\Theta(T/2 - t).$$

From here

$$\text{Det}\,\mathbf{M} = e^{-2igT - iwT/2};$$
$$\text{Tr}\,\mathbf{M} = 2\cos(gT - wT/4)e^{-igT - iwT/4}. \quad (9)$$

In this case there is no instability at all, neither for the reactive, nor for the resistive dampers.

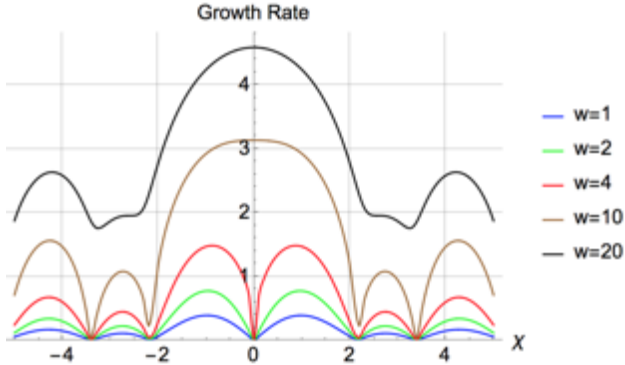

Fig. 4: Growth rates in units of 1/T for the round chamber and zero gain.

## Nonzero chromaticity

Chromaticity is taken into account by adding to the right hand sides of the equations of motion the chromatic terms $\pm i\omega_s \chi \cos(\omega_s t)$, where $\chi$ is the chromatic factor known as the head-tail phase [Chao]. Above the transition, for the leading particle the sign is +, and for the trailing one it is the opposite. With the addition of these chromatic factors, the two-particle problem can be numerically solved. The results for zero gain are presented in Figs. 4-6. Note that all the systems are unstable except for some special values of the chromaticity below certain intensity thresholds. As we may remember, the TMCI thresholds are almost identical for the round and vertical cases, while for the horizontal case the system is stable at this chromaticity for any wake.

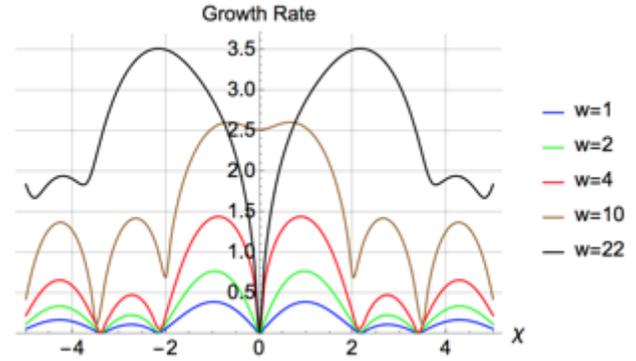

Fig. 5: Same for the vertical case. Note that stability at zero chromaticity is re-established at $w \geq 22$ (mode decoupling).

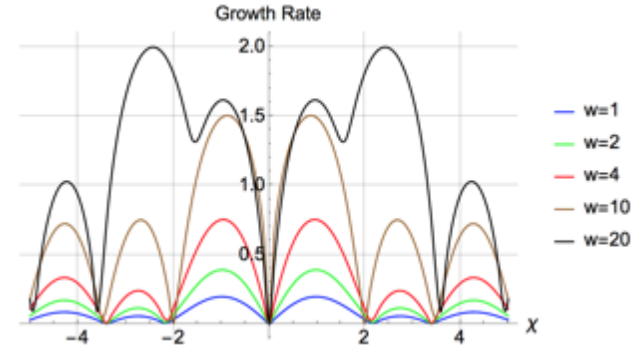

Fig. 6: Same for the horizontal.

Now let's see the effectiveness of the resistive and reactive dampers for nonzero chromaticity; typical examples are shown in Figs.7 and 8.

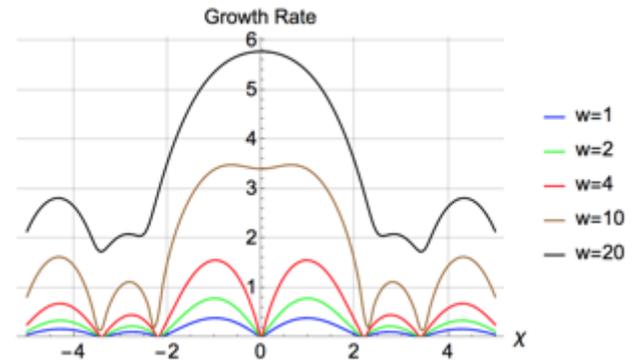

Fig. 7: Round chamber, reactive damper, $gT = 1$.

While the growth rates are reduced at low wakes, they are never zero, as it is the case for the resistive damper, see Fig. 9 and compare Fig. 10a with 10b.

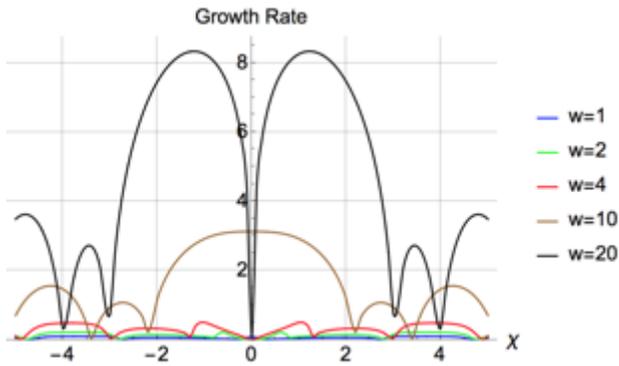

Fig. 8: Same as Fig. 7, but $gT=10$.

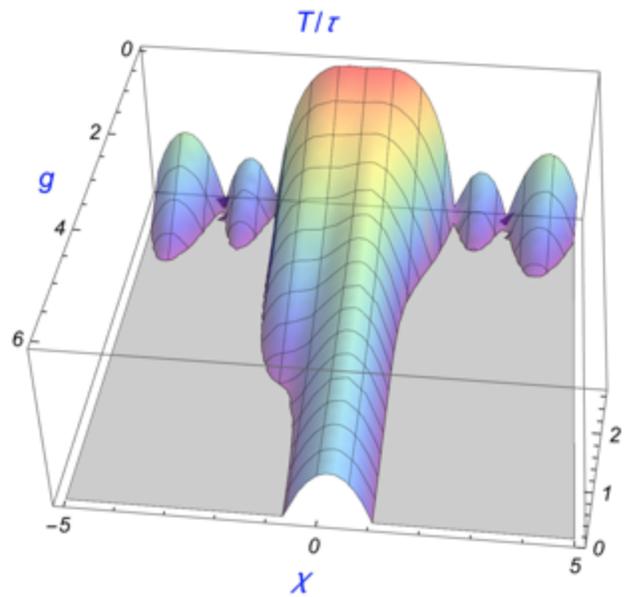

Fig. 10a: Growth rate for the round chamber and resistive damper for $w=8$ (twice above the TMCI threshold). Seas of stability are huge.

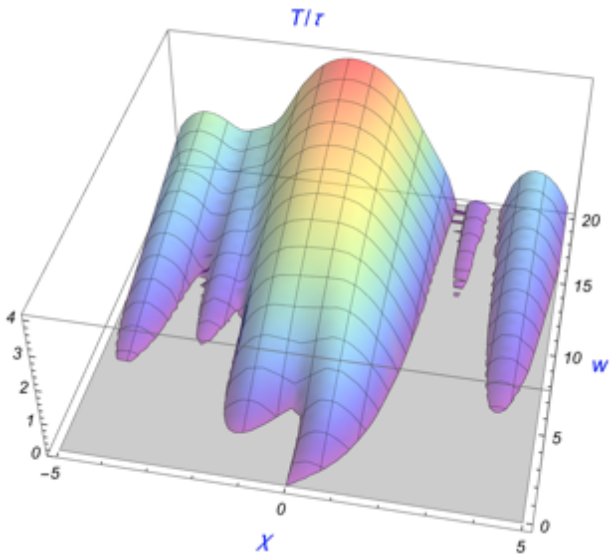

Fig. 9: Growth rates for the round chamber and resistive damper with the gain $gT=1$.

A special interest of this paper is the character of stability areas in the gain-chromaticity plane: they can be 1D or 2D, small or large, open or closed. To characterize them, we will use geographic terminology, calling these areas as rivers (1D), lakes (closed 2D), bays and fjords (semi-open 2D) and seas (huge open 2D).

Mutual comparison of reactive and resistive dampers at nonzero chromaticity shows an advantage of the latter, which provides seas of stability for intensity higher than the TMCI threshold. However, a question can be asked about the validity of this conclusion. In reality, there are many more bunch modes than the two-particle model allows and a variety of wake functions; thus, the real situations can be suspected to be very different. More realistic analysis is suggested in the following sections.

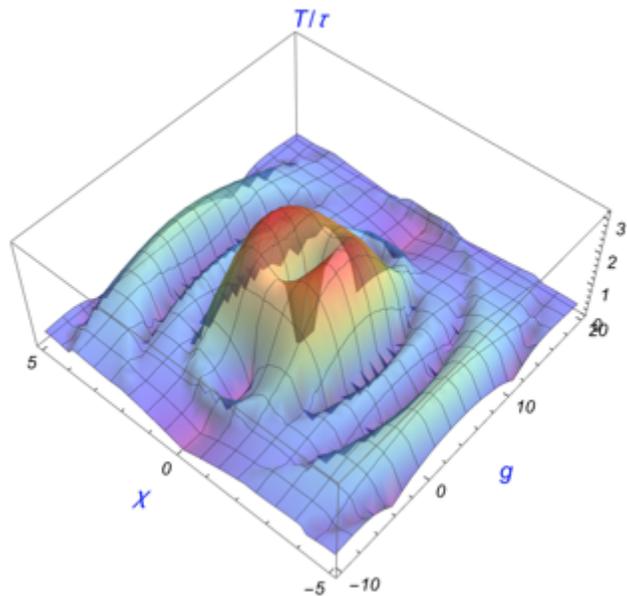

Fig. 10b: Same for the reactive damper. There are no seas or lakes of stability, only rivers.

## NHT WITH BROADBAND IMPEDANCE

Nested Head–Tail (NHT) is a Vlasov Solver for transverse oscillations of bunched beams [NHT] that allows to take into the account azimuthal, radial and multibunch degrees of freedom, influenced by wake fields, feedback and Landau damping. In this section, we discuss the main features of single-bunch instabilities for broadband impedance, taking the ring and bunch parameters of the Advanced Photon Source (APS) of

Argonne National Laboratory, a storage ring of 1.1km circumference and electron beam energy of 7GeV [APS]. We will assume the synchrotron tune $Q_s = \omega_s/\omega_0 = 0.008$, rms bunch length $\sigma_z = 1.5\,\text{cm}$, and rms momentum spread $\delta p/p = 0.001$. The computations are done for a broadband impedance model

$$Z_\perp(\omega) = \frac{\omega_r}{\omega} \frac{R_r}{1 + iQ_r(\omega_r/\omega - \omega/\omega_r)}, \quad (10)$$

with $Q_r = 1$, $\omega_r/(2\pi) = 3\,\text{GHz}$, and the weighted shunt impedance $R_r \beta = 10\,\text{M}\Omega$, where $\beta$ is the average beta-function. The vacuum chamber is assumed to be round.

For the given beam and impedance, NHT computes the entire beam spectrum; the total number of modes is limited by two modelling factors: first, by a number of radial rings representing the bunch longitudinal distribution, and second, by the truncating azimuthal harmonic. For these calculations, the former was taken to be 5, and the latter was limited by ±10; thus, the total number of intra-bunch modes is $(10+10+1)\cdot 5 = 105$.

Growth rate of the fastest growing mode is presented in Figs. 11 and 12 as a function of beam intensity and reactive damper gain; the chromaticity is zero. The intensity parameter $ImpF = N/N_0$ (impedance factor) is defined as a ratio of the bunch population $N$ to its value $N_0 = 4\cdot 10^{10}$ taken as the nominal. The gain $g$ as well as the growth rate $\text{Im}\,q_*$ are measured in the units of the angular synchrotron frequency $\omega_s$. To compare the NHT results with the corresponding two-particle ones, it has to be taken into account, that, by definition, NHT gain units are $\pi$ times larger than the two-particle ones of the previous section, and the NHT growth rate unit is $2\pi$ times higher than the unit accepted for the two-particle model. When comparing units of the intensity parameters, $ImpF$ of this section with $w$ of the two-particle model, one should take into account that the TMCI threshold for the two-particle model $w_{th} = 4$, while the same value for the impedance factor was computed for the broadband case as $ImpF_{th} = 1.6$ (see Fig. 15). Thus, for the sake of comparison of the two models, we may assume that $ImpF$ unit of this section is 2.5 times larger than that of $w$.

To facilitate this comparison, the scale of independent variables, $g$ and $ImpF$, of Fig. 11 are taken in the correspondence with $g$ and $w$ of Fig. 1. One can see that the two figures are rather similar, both in their dominating, merging double-cone structures and the smaller instability areas around. It is worth noting that the growth rate is not a monotonic function of the gain, neither in its focusing nor defocusing direction. At a small gain, $|g| \lesssim 1$, the positive (focusing) sign allows to double the instability threshold, while the defocusing one may reduce the threshold up to a factor of three. However, a further increase of the gain value makes the situation worse in both directions, up to $g \approx -2.5$ when the instability threshold jumps more than 4 times compared to its zero-gain value of 1.6, saturating there for higher defocusing gain values, as one can see in Fig.12. To reach the same threshold for the focusing damper, gain three times higher is needed, while the saturation threshold for the focusing sign is only ~20% higher than for the defocusing one.

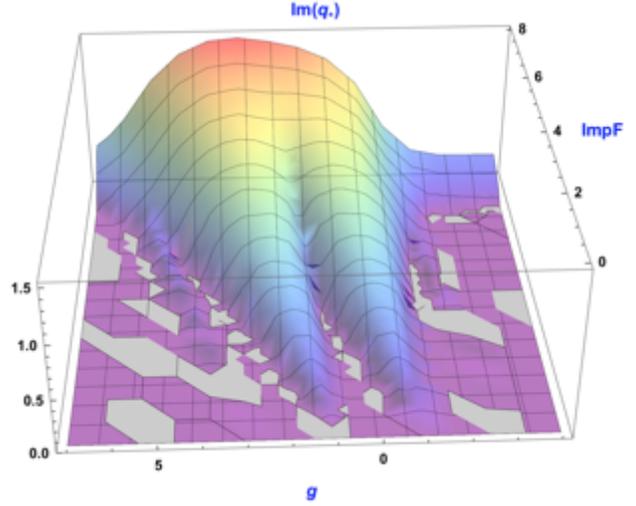

Fig. 11: NHT result for the growth rate $\text{Im}\,q_* = (\omega_s \tau)^{-1}$ versus gain $g$ and intensity parameter $ImpF = N/N_0$ for reactive damper and zero chromaticity. Gain is in units of $\omega_s$; its positive sign corresponds to focusing.

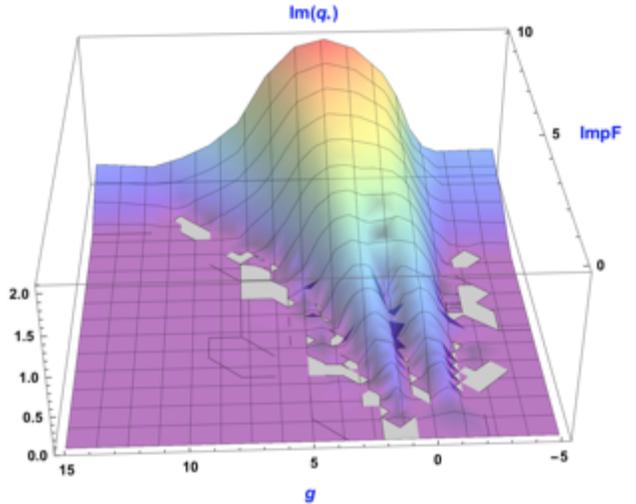

Fig. 12: Same, for a larger range of the variables

As we have seen with the two-particle model, the reactive damper is not very effective for non-zero chromaticity. This conclusion is confirmed by Figs. 13 and 14: outside of a very narrow range of chromaticity, the ravine around zero, the reactive damper is insignificant.

Before going into details of the resistive damper, it is instructive to see the growth rate versus intensity and chromaticity for the no-damper case as it is shown in Fig.15. Similar plots for the resistive case with $g=1$ and $g=10$ are presented in Figs. 16 and 17. Figure 18 shows how the growth rate depends on the chromaticity for various resistive gains, to compare with the similar results for the reactive damper presented in Fig.14. Figure 19 demonstrates that at a high resistive gain and proper chromaticity, the threshold saturates approximately at a four times higher value than the TMCI threshold.

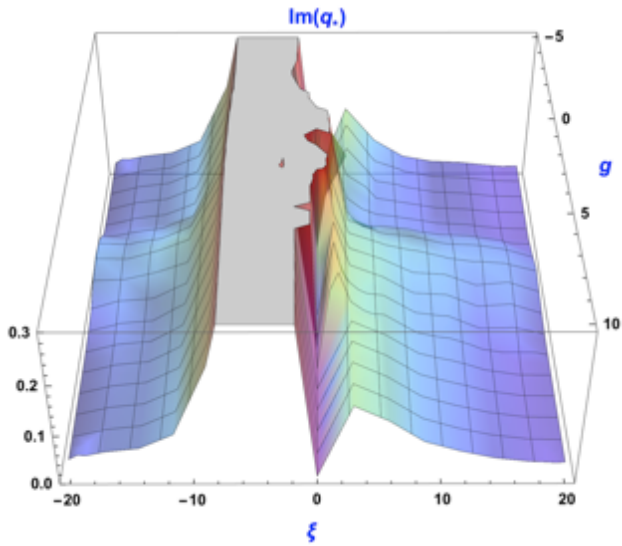

Fig. 13: The growth rate versus chromaticity and reactive gain for intensity twice exceeding zero-gain zero-chromaticity TMCI threshold, i.e. for $ImpF = 3.2$.

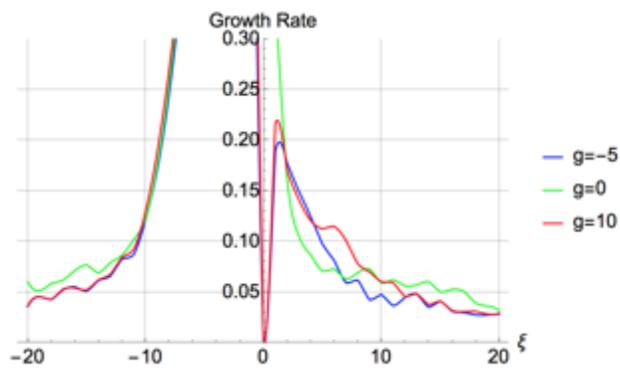

Fig. 14: Same, for three selected gains.

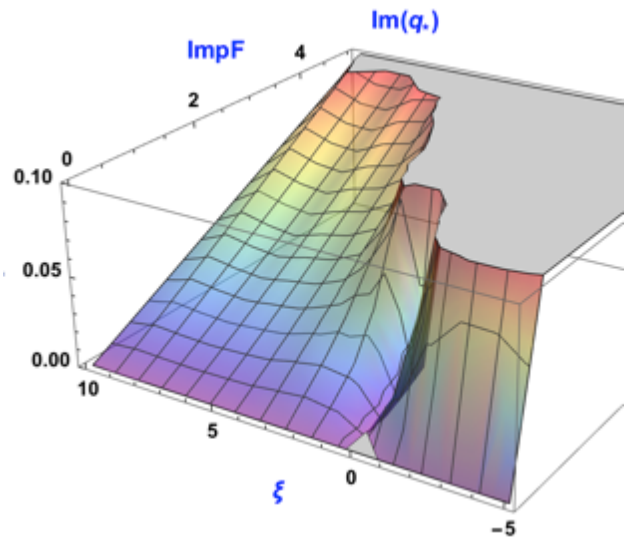

Fig. 15: Growth rate versus intensity and chromaticity; the damper is off.

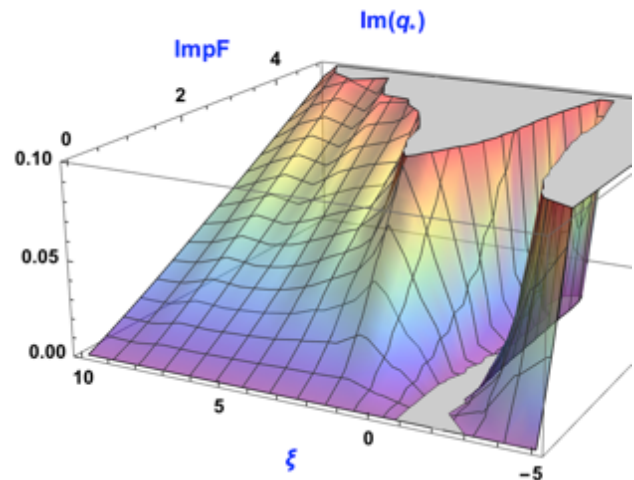

Fig. 16: Same for resistive $g=1$.

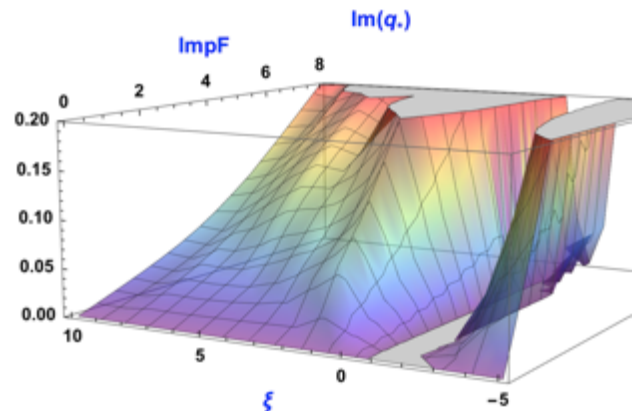

Fig. 17: Same for resistive $g=10$.

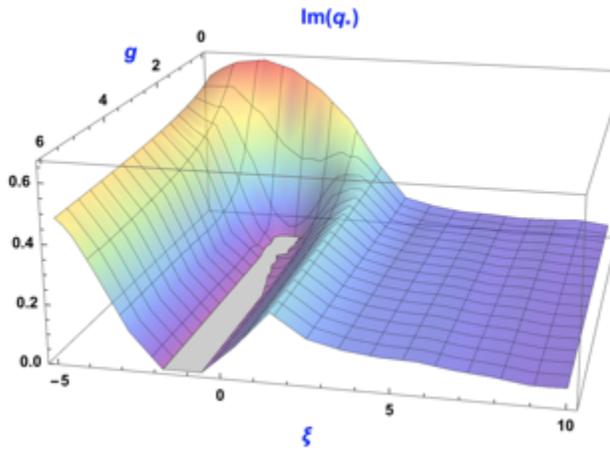

Fig. 18: Same as Fig.14, but for the resistive damper. Note the fjord of stability.

For the reactive damper with any gain, the growth rate can be zero only at zero chromaticity. Contrary to that, for the resistive damper there is a fjord of stability, as one can see in Figs. 16-18. Above the transition energy, this area typically corresponds to small negative values of the rms head-tail phase, centered at

$$\chi_{rms} = \frac{\xi(\delta p/p)_{rms}}{Q_s} \simeq -(0.1 \div 0.2). \quad (11)$$

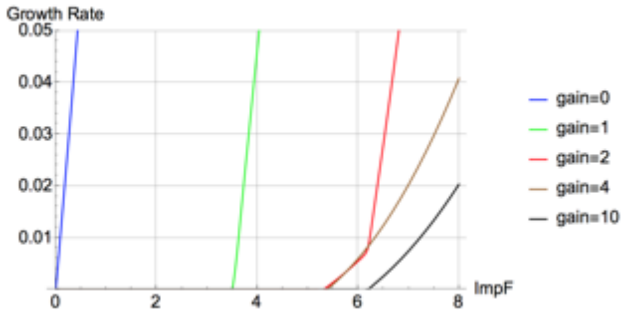

Fig.19: The growth rate for $\xi = -1$ vs. intensity for selected resistive gains. At high gains, the threshold saturates approximately at four times higher value than its damper-off value (6.4:1.6).

The reason for this was in fact explained in Ref. [Myers2]. At low head-tail phase and below TMCI threshold, impedance makes all the modes stable except the zeroth one, corresponding to an almost rigid bunch motion. Since the zeroth mode is perfectly seen by the damper, the resistive feedback provides its damping almost entirely to it. All other modes are poorly seen by the damper at small chromaticity, but there is no need in that since they decay due to impedance. Thus, at low and negative head-tail phase and below the TMCI threshold, the resistive damper stabilizes the only unstable mode and almost does not influence stability of other modes, which are already stable. How far above the TMCI threshold this area of stability may exist is one of the questions of this paper.

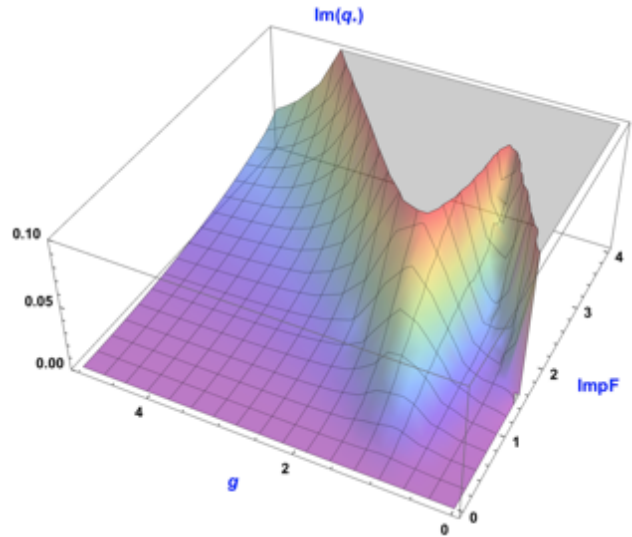

Fig. 20: Growth rate for zero chromaticity and almost reactive damper, which phase declines towards resistive by 18°.

Figures 20 and 21 demonstrate sensitivity of effectiveness of the reactive damper to its small phase variation. These figures show the growth rate versus intensity and gain for zero chromaticity and an almost reactive damper, when its phase declines to the resistive direction by 18°.

Let's imagine, for example, that common action of radiation and Landau damping provides damping rate 0.02, and that available gain cannot be higher than 3.0. Then, as we can see from Fig. 21, this feedback allows increasing the intensity threshold at best by 25%, from 1.6 to 2.0. If the gain deviates from its optimal value in one or another direction, the benefit from the feedback would be even smaller. In this respect, the resistive damper is much more robust also, as one can see from Figs. 22 and 23, where 50% phase deviation towards the reactive one creates almost no effect.

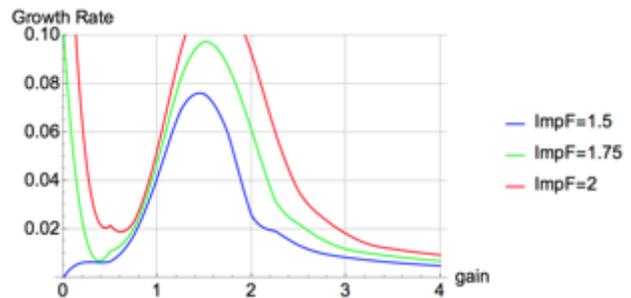

Fig. 21: Same as the previous figure, for selected intensities.

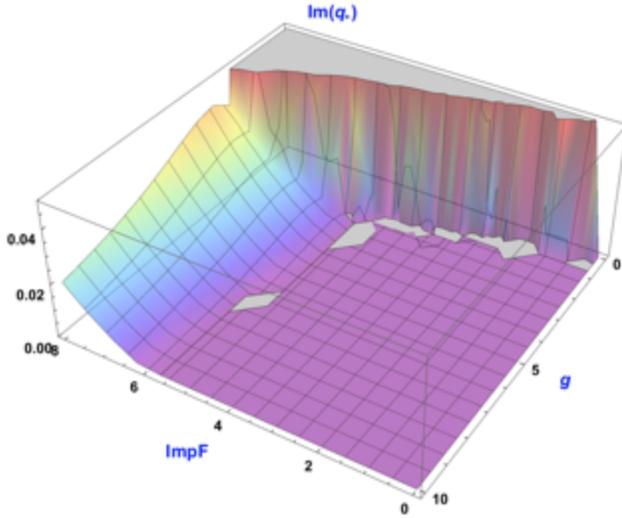

Fig. 22: Growth rate for the resistive feedback and chromaticity $\xi = -1$.

Thus, for the single bunch and the broadband impedance we may conclude about definite advantage of the resistive damper over reactive one. While in both cases the instability threshold, in principle, could be increased up to 4-5 times, tolerance to the offsets of chromaticity and the feedback phase is much better for the former than for the latter.

In the following section we will see how different are the results for the LHC impedance model and how significant can the coupled-bunch contribution be.

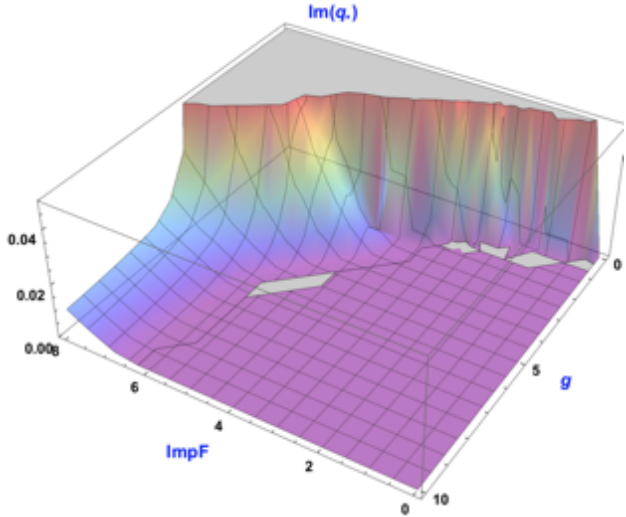

Fig. 23: Same for the feedback phase π/4, i.e. 50% resistive and 50% reactive. Comparison with the previous figure shows how robust the resistive damper is against the phase variations at the proper chromaticity.

## NHT FOR LHC

Transverse instabilities of the LHC beams have been studied with the NHT code in Ref. [NHT] for the Run I parameters. In this section that is reworked with new details for the Run II beam with the energy of 6.5TeV, the bunch separation of 25ns, the synchrotron tune $Q_s = 2.1 \cdot 10^{-3}$, the rms length of a Gaussian bunch $\sigma_z = 7.5$ cm, the nominal bunch population $N_0 = 2.2 \cdot 10^{11}$, and with the same resistive-wall-like impedances [Nicolas].

Figure 24 shows the highest growth rate for a single bunch and no feedback. The TMCI threshold is at $ImpF = 2.4$. Figure 25 demonstrates a decent lake of stability for the resistive damper and the full 25ns beam, with the impedance factor $ImpF = 1.5$, or 62.5% of the TMCI threshold. Note a difference with Fig.18: while for the LHC impedance its area of stability is a lake, for the broadband one it is a fjord. Figure 26 demonstrates one more specific feature of the LHC: the lake of stability disappears almost at the TMCI threshold, $ImpF = 2.4$. Thus, by itself the resistive damper cannot increase the instability threshold for the LHC impedance, even for the single bunch. Variation of the damper phase does not help much: for intermediate resistive-reactive phases the lake disappearance threshold could be increased only by ~15%.

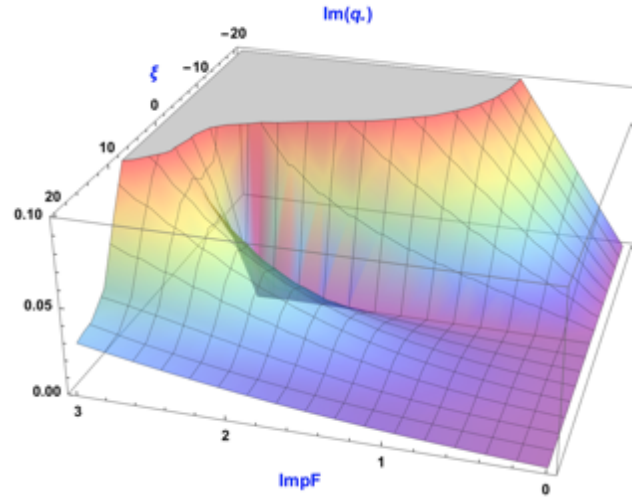

Fig. 24: Growth rate for a single bunch LHC beam and no feedback. The TMCI threshold is at $ImpF = 2.4$

Figure 27 makes clear that the reactive damper, instead, is almost as effective for the LHC as it is for the broadband case: operated at its proper zero chromaticity, for the single bunch it allows to increase the instability threshold more than three times. However, the reactive damper helps very little for the suppression of coupled-bunch instabilities, which all are maximally powerful at zero chromaticity. Thus, for the LHC, with its huge number of bunches, the reactive feedback would not be reasonable. Sufficiently below the TMCI threshold, when the lake is wide enough, the resistive damper tuned to the lake presents an attractive option. Near and above this threshold the only reasonable option for the LHC is to work at the high chromaticity valley of slow instabilities, relying on Landau damping for the suppression of these relatively slow instabilities that remain there when the damper effect is saturated, see Fig. 26, 26a, 26b.

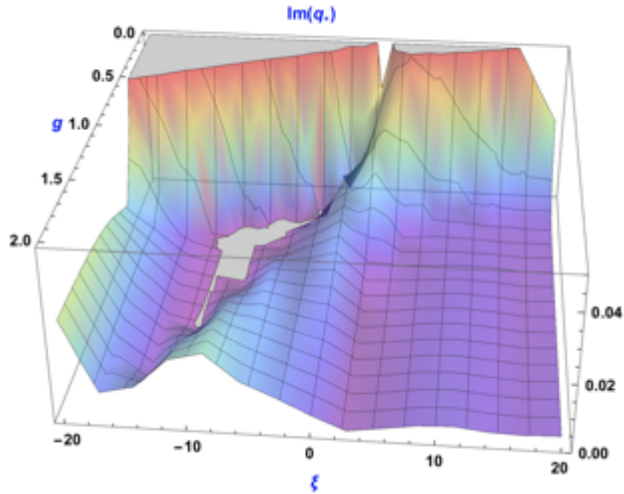

Fig. 25: Growth rate for 25ns LHC beam with the resistive feedback and $ImpF = 1.5$. Coupled bunch interaction is included. Note the lake of stability; for the multi-bunch regime, the lake is limited by $ImpF = 1.7$.

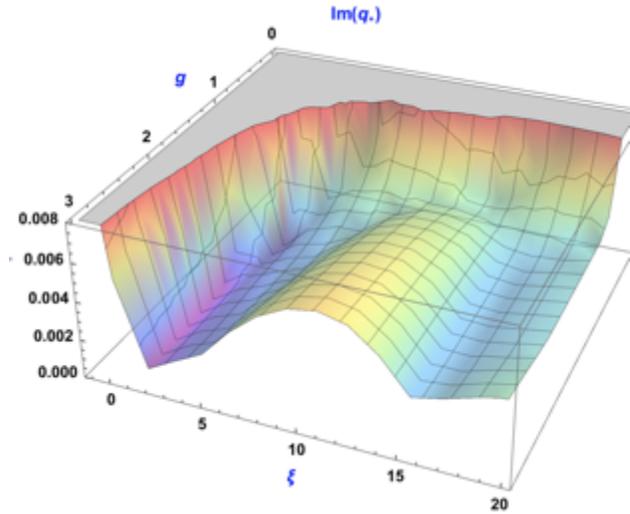

Fig. 26a: The same damper phase and bunch intensity for the full 25ns beam.

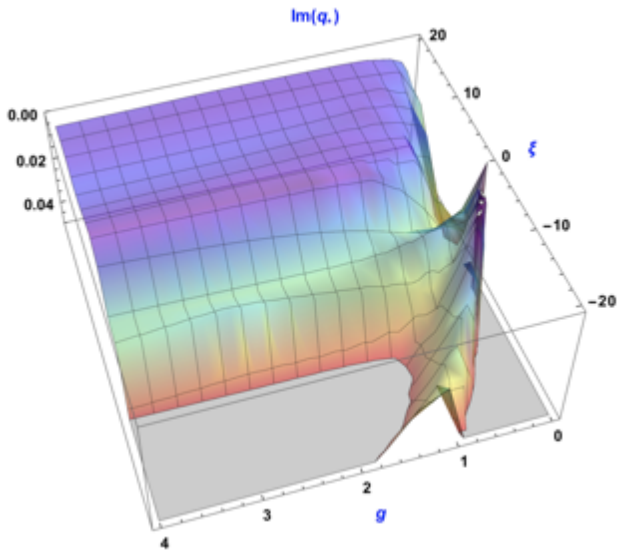

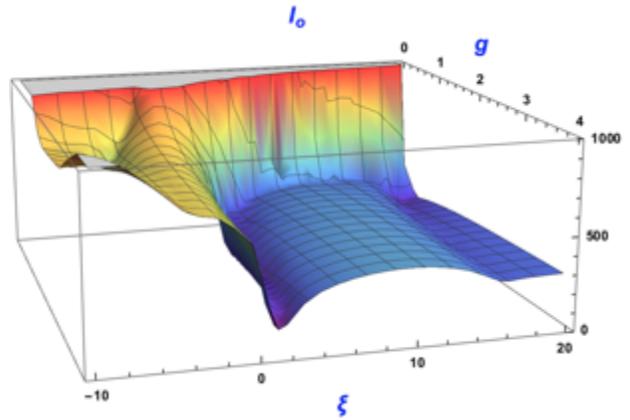

Fig. 26b: Threshold current of the Landau octupoles, in Amperes, for the same case as Fig. 26a, computed according to Ref. [NHT].

Fig. 26: Single bunch growth rate for the resistive damper and TMCI threshold intensity $ImpF = 2.4$. Note that the lake of stability (shown upside down) almost vanished.

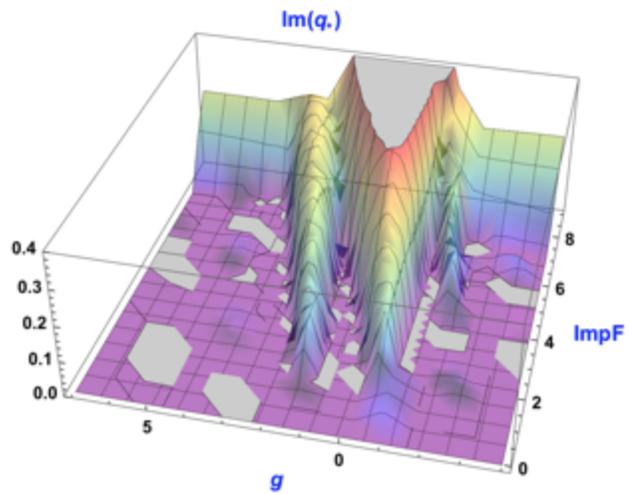

Fig. 27: Growth rate for the single LHC bunch, reactive damper and zero chromaticity. Compare with Fig. 11 for the broadband impedance and Fig.1 for the two-particle model.

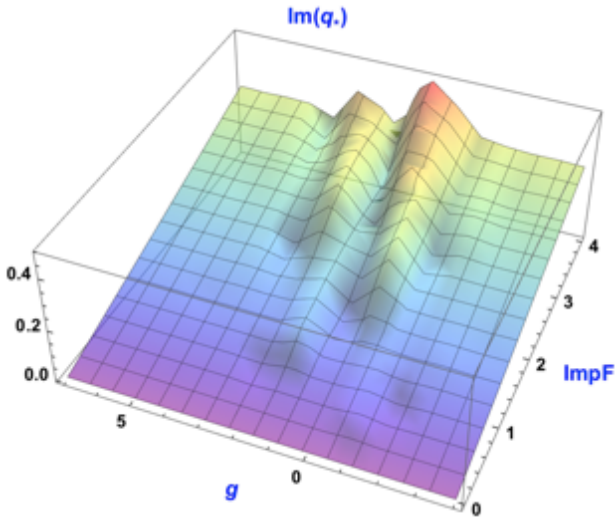

Fig. 28: Same as above, but for the full 25ns beam. The average linear slope towards higher intensity reflects contribution of the coupled-bunch motion. It can be approximated as $\mathrm{Im}\, q_* \approx 0.075\, ImpF$.

## FEEDBACKS AND LANDAU DAMPING

Generally speaking, there are three factors, which may contribute to beam stability: radiation, Landau damping, and feedbacks. The first of them is efficient only for electron beams; it is determined by the beam orbit, focusing, and by sizes of a vacuum chamber shielding coherent synchrotron radiation. This damping is independent of feedbacks, and can be added separately to the total sum of the stabilizing factors. Landau damping is a mechanism of dissipation of a collective mode due to a transfer of its energy to incoherent degrees of freedom of individual particles that happened to be in resonance with this mode. Landau damping is determined by the phase space density of the resonance particles, i.e. both by the separation between the coherent tune and the centre of the incoherent spectrum as well as by the tails of the incoherent spectrum. If the beam is sufficiently relativistic, the space charge effects can be neglected. In such a case, which is the only one considered in this paper, the collective spectrum is determined by the wakes and feedbacks, while the incoherent one is a function of the optics' nonlinearity. Thus, since feedbacks play a role in shaping of beam collective modes, they modify Landau damping also.

With the exception of extremely long bunches or very broadband feedbacks, typical bunch-by-bunch dampers react only on the bunch centroid, kicking the bunch as a whole. As a result, for sufficiently high resistive gain, the bunch center of mass is blocked, while all other possibilities of the bunch motion are not affected by the damper. For round vacuum chambers, as well as for the vertical direction in flat chambers, tunes of modes with significant motion of the center of mass are shifted down for typical wakes [Chao]. Since these center-of-mass dominated modes are normally most unstable, one should expect a certain asymmetry of the modes on the complex tune shift plane. First, with the damper off, this chart of unstable modes should be dominated by the left-hand-sided, or by the negative tune-shifted. When a significant gain is applied, the left-hand-sided modes should be significantly suppressed, while the right-hand-sided, if there are such, most likely should not improve, and might even become worse. That sort of behavior of the chart of unstable modes can be seen in Figs. 29 and 30, for the LHC and broadband impedance respectfully.

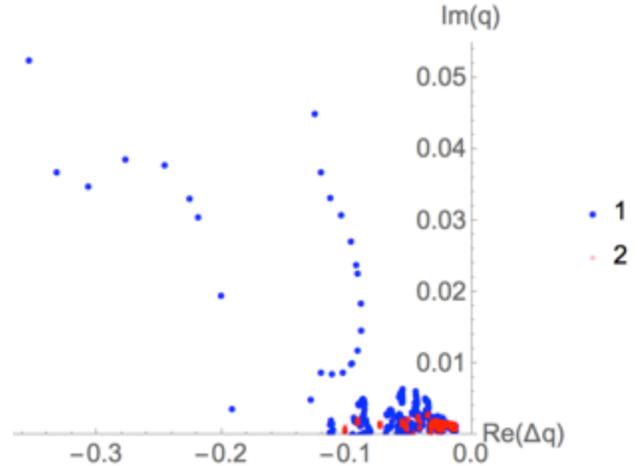

Fig. 29: Tune shifts of unstable modes for the full 25ns LHC beam at chromaticity $\xi = 18$, $ImpF = 2$, damper off (blue, 1), and with resistive gain $g = 1.4$ (red, 2). Both with and without damper, there are no unstable modes with positive tune shifts. Seventeen representative coupled-bunch mode numbers are depicted.

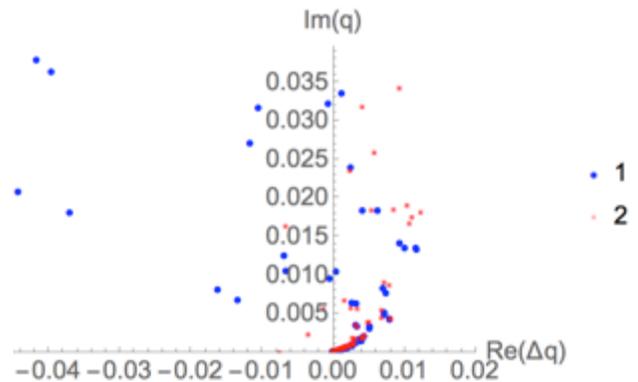

Fig. 30: Tune shifts of unstable modes for the APS single bunch broadband impedance model at chromaticity $\xi = 10$, $ImpF = 2$, damper off (blue, 1), and with resistive gain $g = 1.4$ (red, 2).

We see here a pronounced dependence of the asymmetry on a sort of impedance. With the damper off, both Fig. 29 and Fig. 30 dominate by the left-hand-side modes. However, when it is on, one of them remains to be left-hand sided, while another becomes right-hand-sided. This asymmetry is especially important for electron machines where one of the emittances, the horizontal one, is much higher than another, the vertical. Due to that, transverse nonlinearity makes the stability diagram one-sided too, scaled by the horizontal emittance only, since the vertical emittance is too small to play a role. That is why, for the electron rings, one has to choose whether the diagram has to be designed as right- or left-hand-sided. The correct answer depends, as we just saw, on the type of impedance. Another approach to this problem of the one-sidedness of the stability diagram of electron beams is to provide electrons with the missing sign of the tune shift by means of the second order chromaticity which sign is made opposite to the one of the horizontal nonlinearity.

## CONCLUSIONS

Possible strategies of beam stabilization by means of a feedback were considered with three models: two-particle model, NHT broadband impedance model and NHT with the LHC impedance model. Advantages, challenges and limitations for reactive and resistive dampers are formulated. Existence and details of the 2D area of stability in the gain-chromaticity plane is shown to depend on the type of impedance. One-sidedness of the mode tune shifts, as well as stability diagrams is pointed out as a source of instability. Possible solutions for this problem are outlined.

## REFERENCES

[Kohaupt] R.D. Kohaupt, "Simplified Presentation of Head - Tail Turbulence" DESY M-80/19 (1980).

[Talman] R. Talman, "The Influence Of Finite Synchrotron Oscillation Frequency On The Transverse Head-Tail Effect", CERN ISR-TH/81-17 (1981); Nucl. Instr. Meth. **193**, 423 (1982).

[Ruth] R. Ruth, "Reactive Feedback In The Two Particle Model", CERN LEP-TH/83-22 (1983)

[Chao] A. Chao, "Physics of Collective Beam Instabilities in High Energy Accelerators", p. 142 (1993).

[NHT] A. Burov, "Nested head-tail Vlasov solver", Phys. Rev. Accel. Beams 17, 021007 (2014).

[Nicolas] N. Mounet, "The LHC Transverse Coupled-Bunch Instability" CERN-THESIS-2012-055 (2012).

[Myers] S. Myers, "Stabilization of the Fast Head-Tail Instability by Feedback" Proc. PAC'87, p. 503.

[APS] M. Borland, G. Decker, L. Emery, W. Guo, K. Harkay, V. Sajaev, C.-Y. Yao, "APS Storage Ring Parameters", 2010.

[Myers2] L. Amaudon, S. Myers, R. Olsen and E. Peschardt "Transverse Feedback and LEP Performance", Proc. EPAC'92, p.66.